# Framework of SQL Injection Attack

Neha Patwari[1], Parvati Bhurani [2]

## Abstract

With the changing demographics of globalization, the emergence and prevalence of web application have acquired a central and pivotal role in the domains of technology and advancements. It thus becomes imperative to probe deeply into the architecture, significance and different facets of usages. Web applications enclose the functioning between a user and the services provided by the server, which contains a database as its backend. The user can access the required information through sending a request in the form of text to the web server, which is interpreted by the server side script to construct an SQL. The query is sent to the database which responds in order to generate an HTML page that is sent back to the user. Since the functioning of web application is a dynamic and complicated matter, certain threats to the database security have been registered. One such alarming threat is the prevalence of SQL Injection Attack. Hence a dynamic algorithm is given in this paper for preventing SQL Injection Attacks which is based on context free grammars and compiler parsing techniques. The paper attempts to present the notation of a

SQLI Prevent Parser for the prevention of SQL Injection Attacks. This Parser determines the structure of queries and compares whether the queries are functionally equivalent or not. This parser has been used on a sample web application and the results have come out to be positive majors to prevent SQL Injection Attacks.

## I Introduction

There has been a rapid advancement in information technology as a result of the widespread use of the internet since the past few years. The common man today uses the internet with a number of purposes such as to be used in the field of education, for money transactions and other countless activities. Though there is also an inherent risk in the frequent use of the internet as found in transferring some money from one bank account to another or in the confidential database of the companies. The secure websites stores the highly sensitive information along with non-critical data in their database systems in such a way that the Owner of the information is able to access it quickly while attackers of the unauthorized users are blocked in their attempts to have access to the information.

Hence we have to understand the architecture of web application; a web application accepts requests from users in order to gather information from a database. It is assumed by database that the input is correct and thus uses it to access the database by creating an SQL. These web applications become virtually prone to SQL injection attacks since these do not check the validity of the user queries before submitting them to gather the data. For example, attackers pretending as genuine user utilize maliciously created input text which contains SQL instructions in order to produce SQL queries on the web application back-end. In case web application processes the query, the accepted malicious query may breach security net of the underscored database. As a consequence of the query there occurs an improper functioning of the database parser which results in the release of the sensitive information [1].

In order to have access to the sensitive information from the database a general break-in strategy is to first create a query which will corrupt the functioning of the database parser, and forward the application of this query to the targeted database. This type of approach in order to have access to the private information is known as SQL injection. Now SQL injection has become a common occurrence due to the easy access of the database via the internet. It is equally necessary to have a deep understanding of the types of communication which occurs during a particular session in between a web application and a user in order to have a better understanding of SQL injection.

## II Overview of SQL injection

A web application is one through which a user can access the services provided by the web server while working on a client machine, which contains a database for example an online email id. The user enters a login name and password to access the email account. As he presses the submit button a URL is created and is sent to the web server. The server side of the script interprets the user input due to which a dynamic SQL query is created. It is submitted to the database and HTML pages are generated in response to the query which is sent back to the user. A particular section of the database query code is submitted by the malicious attackers to the server, while responding with the corresponding result some sensitive information is disclosed by the server. This is categorized as SQL injection attack. A SQL injection attack contains injection of a SQL query to the application through the input data from the client. If successful SQL injection can read and modify the data in the database (Insert/Update/Delete), it means that an SQL injection attack takes place. When the intended effect of an SQL query is modified by an attacker on inserting new SQL keywords of operators into the query, following are the qualities of SQL injection attacks:

i) Threat Modelling  ii) Attack Intent  iii) Assets

## III. Working of SQL Injection

The concept driving a SQL injection is simple above all attacks like these can be executed and mastered with ease. To exploit the SQL injection weakness the basic requirement for the attacker is to identify the working of the web application. A malicious SQL command can be inserted carefully into the content of the criteria empowering the attacker to trick the web application so that a malicious query can be forwarded to the database.

e.g. the LOGIN FORM which accepts the username and password from the login. The input in the field ("name" and "Password") is directly used to create the SQL Query like:

**SELECT * FROM login WHERE name = 'name' AND password = 'password';**
Now, let the user input the correct name ="Administrator" and Password="admin". The query will become:

**SELECT * FROM login WHERE name = 'Adminstrator' AND password = 'admin';**

This will function without any problem. In case the user supplied some vulnerable string of code then that will empower the attacker to by-pass the authentication and create an SQL Injection so that he finds out the relevant information from the database. i.e., if user inputs username= 'OR 1=1- -then the query will be forwarded as :

**SELECT * FROM login WHERE name = ' 'OR 1=1–'AND password = ' ';**

It will work as specified below:
The input data is being used in the WHERE clause. Since the application is not actually concerned about the query simply tailoring a string, user has converted a single-component WHERE clause into a two-component clause, and this makes it certain that the 1 = 1 clause will be true notwithstanding the fact that what the first clause is. The query emphasize that "Select everything from the table login if the name equals "nothing" Or 1=1, ignores anything after the comment.
' : Is used to close the user input field.
OR : The SQL query will be continued to get the process as equal to what proceeds before OR what follows.
1=1 : A statement which is all time true.
– : Discards the rest of the lines in order to stop further processing.
Noticing that 1 will always equal 1, the server has been virtually duped as the statement received is true and this empowers the attacker to have additional access. The code which relates to the password input field is not run by the server and therefore does not use it [2].

## IV. Types of SQL Injection Attacks

Divergent types of advanced and powerful techniques have been developed by attackers over the past several years which empower attackers to exploit SQL injection vulnerabilities. These techniques are much advanced than the generic SQL injection attacks examples and derive the benefits from sophisticated SQL designs. These threats must be taken into account while working on the development of SQL injection attack problems.

An SQL Injection Attacks proneness can be exploited by the attacker once he has detected the input source, for this purpose the attacker can utilize various types of techniques. As per the type and extent of the proneness the attack can lead to crashing the database, collecting the relevant information regarding the tables in the database. Given below is a synopsis of the main techniques of performing SQL injection attacks.

An isolated attack is not a general phenomenon instead a combination of attacks either simultaneously or sequentially used as per the desired target of the attacker.

## First Order Attacks

In some attacks the desired result is immediately received by the attacker. This may be due to the direct response by the application with which they are interacting or may be via some other response mechanism for example E-mail.

All type of attacks which is mentioned below, if performed directly in text field and provides important information or data, from the response then such type of attack is called First Order Attack or Direct injection.

In the case of direct injection the SQL query will use each argument submitted as such without any modification. For example attempt to take parameter's legitimate value and appending a space along with the word "OR" with it. In case if an error is generated by this, a direct injection is possible.

First order is basically performed by SELECT query which is used in application for retrieving information.

### Tautologies Queries

**Attack Intent:** Bypassing authentication, retrieving data, identifying inject able parameters.

**Description:** The normal aim of this type of attack is to inject code that may be in one or more conditional statements due to which the statements are always evaluated as to be true. The results of this type of attack take place due to the way in which the application uses the outcome of the query. The most common purpose is to skip authentication route and extract data. An attacker exploits an injectable area under this type of injection which is utilized in a query's WHERE conditional. The transformation of the conditional into a tautology results in returning all the rows in the database table being targeted by the query. For the attack to be fruitful the code must either display all of the returned records or must perform some action so that at least one record is returned.

**Ex:** Let there be an input form with the fields "name" and "password". Using this user can login in web application. The given below PHP code for the application server, created by a web application developer has inherent weakness for SQL injection attack:

1. $connection=mysql_connect();
2. mysql_select_ db("sample");
3. $user=$HTT_GET_VARS['name'];
4. $pass=$HTTP_GET_VARS['password'];
5. $query="select * from login_table_ll where name='$ u_user1_name 'and password ='$p_pass1_name'";
6. $result=mysql_query($query);
7. if (mysql_num_rows($result)==1) echo "Authorized" else echo "authorization failed";

User data created in the form of a web are assigned to variables "u_user1'_name' and "p_pass1_name" and then utilized to produce the SQL statement.

Query (i) given below is generated after entering valid name 'adminstrator1_ad'and valid password 'admin1_ad'by genuine user.
Query = "select * from login_table_llwhere name='adminstrator1_ad'and password='admin1_ad'";————(i)

If an attacker writes: 'or 1=1–'in the name field (the input entered for the other fields are impertinent) leaving the password field empty, the structure of the SQL query will be changed.
Query (ii) given below is generated with SQL injection by the attacker.
Query = "select * fromlogin_table_ll where name=''or 1=1 –'and password=''————(ii)
The complete WHERE clause is transformed into a tautology by the code injected in the conditional('OR 1=1–). The conditional is used by the database as the basis in order to evaluate each row and to decide which is to be returned to the application. As the conditional being a tautology, the query evaluation is true for each row in the table and so all of them are returned [3].

*Illegal/Logically Incorrect Queries*

**Attack Intent:** Retrieving data, identifying inject able parameters, performing database finger-printing.
**Description:** This category of attack allows to collect the relevant information as per the type and structure of the back-end database of a Web application.

The main aim of this attack is to gather information for further attacks and is treated as a preliminary step. These attacks pinpoint a weakness due to which the application servers returns the default error page which often contains over description. The vulnerable or inject able parameters can be revealed to the attacker due to the simple fact that error messages are being generated.

The additional error information which was fundamentally aimed at assisting the programmer to repair or correct their application further empowers the attacker to access information related to the schema of the back-end database. During working on this type of attack, he tries to inject statements which can result in syntax error, type conversion or could create logical error into the database. The injectable parameters can be detected by using the syntax errors. The deduction of the data types of certain columns or the seperation of the data can be done by using the type errors. The names of the tables and columns causing the errors can often be revealed by logical errors.

**Example:** In case the syntax error consists of a parentheses in the cited string (for example SQL Server message used in the illustration given below) or a message is generated which clearly mentions about missing parentheses.

A parentheses must be added to the bad value part of the injection, and one to the WHERE clause. In few cases two or more parentheses may be required. Here's the code:

mySQL= " SELECT Last_ name1_l, First_name1_f, Title_ t1_t, Notes_n1_n FROMEmployee_ Table1_eWHERE City_ name1_c = (' "& strCity &" ') "

When an attacker inserts " ' " then the query is built as:

"SELECT Last_name1_l, First_name1_f, Title_t1_t, Notes_n1_n FROM Employee_Table1_eWHERE

City_name1_c =(' ')"

Then the error generated is :

Error Type:

Microsoft OLE DB Provider for ODBC Driver [Microsoft][SQL Server ]Unclosed Parentheses mark before the Character String " ' " From the error generated, the attacker knows that here parentheses is used.

Hence, attacker tries to inject the value ') ('UNION SELECT another field FROM another table), thus this query will be forwarded to the server.

SELECT Last_name1_l, First_name1_f, Title_t1_t, Note_n1_n FROM Employee_table1_e WHERE City_name1_c = (' ') ('UNION SELECT another field from another Table ') ;

Through the errors generated, the attacker gets to know a lot of useful data through various steps.

Hence by the use of error messages attacker gets information [4].

### _Union Queries_

**Attack Intent:** Bypassing Authentication, extracting data.

**Description:** In such attacks the weaker parameters are exploited by the attacker with a view to transform the data set returned for a specific query.

This technique allows the attacker to form the application, giving back data from a table not from the one which intended by the developer but from another unintended table.

The attacker performs it by introducing a statement in the way:' UNION SELECT < remaining of injected query >.

Since the second/injected query is totally controlled by the attackers, this query can used by them in order to retrieve information from a particular table. This attack results in the form of a dataset from the database which is the collective result of the original query and the injected query.

**Example:** Referring to the running example, an attacker could introduce the text " 'UNION SELECT card_no1_c from Debit_Card1_d where accountNo1_a=100–" into the login1_area field, leading to the generation of the following query:

**SELECT bank_accounts FROM users1 WHERE login1_area = ''UNION SELECT card_no1_c from**

**Debit_Card1_d where accountNo1_a = 100 – AND pass='';**

Predicting that there is no login1_area equal to " ", a null set is returned by the first original query, while the data from the "Debit_Card1_d" table is returned from the second query. For account "100" the column "cardNo1_c" would be returned by the database in this case. The result obtained from these queries combines and returns them to the application [3].

**Second Order Attacks**

In this type of attack when the malicious code is injected into the web based application instead of being immediately executed it is stored by the web application i.e. it is first stored in the database to be retrieved, rendered or executed by the victim. This category of attack happens because of the notion that when the data is contained in the database, it is often supposed to be clean and need not be checked again. While due to the frequent use of the data in the queries, it is still able harm the web application. This type of attack happens in case where the filtration process is skipped during the process of data insertion in search page. We should apply filtration for special characters before storing data in databases, which no special characters are allowed for inserting in databases. It is inherently performed by INSERT basics which are used in application. INSERT keyword is used to add information in the database. In case of web application this keyword is used for user registrations, bulletin boards inclusion, adding items to shopping carts, etc. While trying to INSERT injection it could result in the flooding of the rows in the database having single quotes and SQL keywords. As per the at tentativeness of the administrator it can be evaluated that what is to be done with the information. For example the user is on a site on which user registration of some kind is allowed.

A format is provided in which the user has to enter name, address, phone number, etc. As the information is submitted in the format a page is generated where this information is displayed along with an option to edit the information. This is what is required by the user. Thus after the process of insertion the required data can be modified and updated. Thus in case some malicious data is inserted in the database by the attacker, the data can be updated as per the desire of the attacker.

Piggybacked Query attack is example of Second Order Attack [5].

## Piggybacked Queries

**Attack Intent:** Inserting or updating data, performing denial of service.
**Description:** In this category of attack, attacker tries to inject queries in the original query. These kinds of attacks in contrast to the other type of attacks instead of modifying the original intended query tries to insert new and distinct query that "piggy-back" on the original query. This results in multiple SQL queries to the database. The initiating query (intended query) is executed as normal while the remaining queries are injected queries, and being executed along with the initiating query. The attack of this category is highly fatal. In case an attacker succeeds in this attack he can virtually insert any sort of SQL command in the additional queries and is able to execute them along with the initiating query. This kind of attack vulnerability is often due to the possession of a database configuration via which multiple statements can be inserted in a single string.

**Example:** If the attacker inputs " '; drop table login ;" into the password field, the application generates the query:
**SELECT * FROM login WHERE name='admin' AND password= ' '; drop table login ;**
As the first query having query delimiter (";") is completed the second query is executed by the database. The effect of the execution of the second query would be to drop table login which may lead to the destruction of the valuable data [6].

## V. Prevention Methodology

The methodology which has been used to prevent the SQL injection attacks is the merging of SQLIPreventParser with the application therefore protecting against any attacks. Firstly SQLI Prevent Parser has been built which is used to determines the structure of the query. Then limitations of the method are identified. Finally, the solution to overcome the problems has been proposed making the system fully efficient.

### *Approach*

The developer built a data structure for the parsed representation of the statement, which is called a parser. For parsing, we require the grammar language of statement. In this method, by parsing two statements and comparing their parser functionality, it leads to conclusion that the two queries are equal. When sql is injected successfully in database query, the parser of the intended SQL query and the resulting SQL query is generated after mismatch of attacker's input.

The SQL Query is:
**SELECT * FROM login WHERE login name=' ' AND password=' ';**
Web applications have SQL injection vulnerabilities because inputs are not sanitized which they use to construct structured output.

If an attacker passes name = ' OR 1=1– as the login name, all login name in the database will be returned and displayed, reason being transformation of entire WHERE clause into a tautology of code injected in the conditional statement('

OR 1=1 –). The conditional used by database to evaluate each row and decide the rows to return to the application. Since conditional is a tautology, query evaluates each row in the table as true and returns all of them to application. The problem is reckoned by taking into consideration its cause:

The detailed information of the program is that the substrings are taken from user input and the substrings are restrained syntactically. The concept is to restrict queries in which the input substring modifies the syntactic structure of the remaining query. Such queries are called SQL injection attacks in the perspective of database back-ends.

The user's intake is visualized by using meta-character displayed as '(|' and '|)'. It allegorizes the commencement and ending of each input string. This meta character follow the string through assignments, concatenations, etc., thus as a query is ready to be transferred to database, it contains matching pair of markers identifying the substrings from input. We should refuse to introduce input substrings from modification of the syntactic structure of the remaining the query. For this grammar for queries as per the standard grammar for SQL queries is build up. In the grammar, the only productions in which '(|' and '|)' occur have the following form:

Non terminal ::= '(|' symbol '|)'

Where symbol is either a terminal or non-terminal

For query to be in the language of this grammar, the substrings surrounded by '(|' and '|)' must be syntactic. A parser generator is used to build a parser for grammar and each query is attempted to be parsed. In case the query is parsed successfully, it meets the syntactic constraints and is legitimate. Conversely, it fails the syntactic constraints and may be a SQL injection attack.

After SQL Prevent Parser is built using the grammar of the output language and plan of action is specified that permitted syntactic forms, it remains on the web server and intercepts generated queries. Each input needs to be propagated in form of some query, notwithstanding the input's source, gets amplified with the meta-characters '(|' and '|)' Then query is generated by the application, which SQLIPreventParser attempts to parse. If a query parses successfully, SQLI Prevent Parser sends it to the database without the meta-character. Otherwise, the query is block out.

## VI. Limitation

This solution can be overcome in either of two ways:

1. If the attacker is somehow able to detect the delimiter used, it would

   require only a slight modification of the query to break this protection.
2. The attacker may simply use a brute force attack to simply try out all possible combinations (to guess the correct delimiter combination).

## VII. Solution

Original solution where we use static delimiter upgraded to circumvent potential security leaks. Hence implementation by dynamically changing the delimiter combination for every variable field and not using the same delimiter blend for two consecutive variable fields or in same field in application. As a result of this modification to the original algorithm, the attacker will have to correctly guess the exact sequence of delimiters used to bypass the parser's security system. Since the delimiter blend will be cycled randomly this will not be easily possible. By this proposed method static delimiter has been made dynamic. This solution makes the parser more secure than before.

Figure 3.3 shows the basic structure of work where the user input is interpreted by the web application. In the web application it has been used the concept of dynamic delimiter so that the attacker is unable to guess the sequence of the delimiter; here even the user has no idea about the sequence of the delimiter.

Hence in the given application the limitation of the static delimiter has been eliminated.

Figure 3.3: Work Architecture.

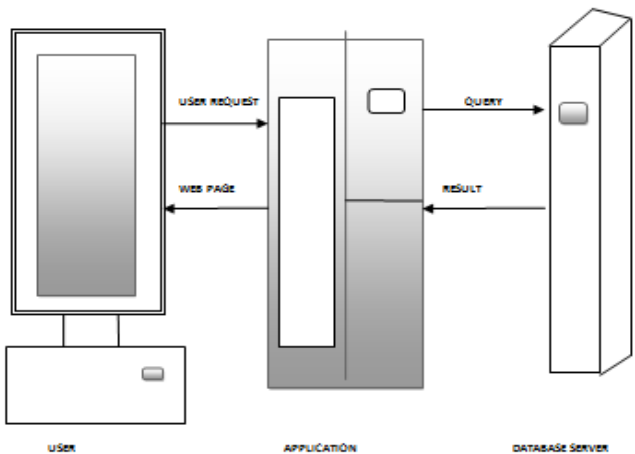

If the user puts any input by using the delimiter for example x|)' OR '(| 1 = 1 , then it will be checked at the application itself and the error is reported here itself. Now from application query is sent to the parser.

The Parser determines the structure of SQL query and input variable. Parser compares that both queries ( means query at the application and the query at the Parser) are functionally equivalent or not. Incase both the queries are functionally equivalent then it reaches the database then response is taken from the database, which is generated as an HTML Page and is send to the user [1].

## VII. Result

This paper presents the first overview of SQL injection attacks in web application. According to the presented paper an effective technique has been developed for preventing SQL injection attacks. The implementation on web application and parser on java CC [7] proved effective under testing. Here have been diligent efforts in applying parser on web application and produces output.
The result of evaluation and test proves that the proposed method is an effective technique to prevent SQL Injection Attacks.

In this work it has been managed to prevent SQL injection attacks through:

• Tautologies Queries
• Union Queries
• Illegal/Logically Incorrect Queries
• Piggybacked Queries
•SQLIPreventParser has been built for SQL constraints

Following are the two goals for future works:
   1. The parser is to be more generalized for maximum number of SQL commands.

2. The technique can be applied to prevent cross-site scripting.